\documentstyle[11pt]{article}
\newfont{\bbbold}{msbm10 scaled \magstep1}

\def\bbR{\mbox{\bbbold R}}

\newfont{\goth}{eufm10 scaled \magstep1}

\def\gi{\mbox{\goth i}}

\def\gn{\mbox{\goth n}}

\def\gp{\mbox{\goth p}}

\def\gs{\mbox{\goth s}}

\def\a{\alpha}
\def\b{\beta}
\def\c{\gamma}\def\C{\Gamma}
\def\d{\delta}
\def\e{\epsilon}

\def\h{\eta}

\def\L{\Lambda}

\def\beq{\begin{equation}}\def\eeq{\end{equation}}
\def\beqa{\begin{eqnarray}}\def\eeqa{\end{eqnarray}}
\def\barr{\begin{array}}\def\earr{\end{array}}

\def\O{\Omega}

\def\xz{\times}

\def\hT{{\hat{T}}}
\def\hR{{\hat{R}}}
\def\hD{{\hat{D}}}

\let\la=\label

\def\bd{\begin{document}}
\def\ed{\end{document}}
\def\ba{\begin{array}}
\def\ea{\end{array}}
\def\bea{\begin{eqnarray}}
\def\eea{\end{eqnarray}}
\def\ft#1#2{{\textstyle{{\scriptstyle #1}\over {\scriptstyle #2}}}}
\def\fft#1#2{{#1 \over #2}}
\newcommand{\be}{\begin{equation}}
\newcommand{\ee}{\end{equation}}

\begin{document}

\begin{titlepage}
\begin{flushright}
King's College/KCL-TH-97-30\\
hep-th/9707184\\
\today
\end{flushright}
\vskip 2cm
\begin{center}
{\Large{\bf {Weyl Superspace}}}
\end{center}
\vskip 1.5cm
\centerline{\bf P.S. Howe,}
\vskip .5cm
\centerline{Dept. of Mathematics,}
\vskip .5cm
\centerline{King's College,  London, UK.}
\vskip 1.5cm

\begin{abstract}

\noindent
It is shown that the equations of motion of eleven-dimensional supergravity
follow from setting the dimension zero components of the superspace torsion
tensor equal to the Dirac matrices. The proof of this assertion is facilitated
by the introduction of a connection taking its values in the Lie algebra of
$CSpin(1,10)=Spin(1,10)\times \bbR^+$. The resulting formulation of
eleven-dimensional supergravity is Weyl covariant but is equivalent to the
usual formulation modulo topological considerations.

\end{abstract}

\end{titlepage}

In view of the current interest in eleven-dimensional supergravity it is useful
to try to gain a better understanding of the superspace geometry of the theory.
In this note we show that this geometry is very tightly constrained to the
extent that the standard choice for the dimension zero torsion is actually
sufficient to put the theory on shell. This may be useful, for example, from
the point of view of trying to find higher order corrections to the theory
which are expected to be present in the conjectured $M$-theory. The analysis of
the consequences of the basic constraint is simplified by the introduction of a
connection which takes its values in the Lie algebra of
$CSpin(1,10)=Spin(1,10)\times \bbR^+$, and we call a superspace with such a
connection Weyl Superspace. The formulation of supergravity we shall arrive at
is therefore locally scale invariant, but in a somewhat trivial way. The
associated scale curvature tensor vanishes and so  one can recover the usual
theory in a suitable gauge, at least locally.

The equations of motion of eleven-dimensional supergravity \cite{cj} were
presented in the superspace formalism some time ago \cite{cf,bh}, and the
constraints have been interpreted in terms of membranes \cite{bst,dhis} and as
integrability conditions in membrane superspace \cite{h}. More recently, the
authors of ref. \cite{cl} carried out an analysis of the geometrical superspace
constraints (i.e. without including the four-form field strength initially) and
showed that the standard dimension zero torsion constraint implies that, at
dimension one-half, there is a single spinor field appearing in the torsion.
They then set this field to zero and recovered the standard on-shell formalism.
In this note we ask what happens if one does not set this spinor field to zero.
It turns out, ignoring topological niceties, that it can be written as the
derivative of a scalar superfield and that this scalar superfield can be
transformed away by a super-Weyl transformation. It is, however, easier to
demonstrate this by modifying the formalism slightly to include a scale
connection in addition to the usual Lorentz connection. In fact, the result
implies that we can summarise the equations of motion of supergravity in eleven
dimensions without introducing a connection at all, and we start off with a
brief discussion of this before reverting to the connection formalism to prove
the main point.

The basic structure we shall study can be called a special superconformal
structure in $(11|32)$-dimensional superspace $M$. Such a structure is a choice
of odd tangent bundle $F$ having rank $(0|32)$ with associated Frobenius tensor
which is maximally non-integrable and invariant under
$CSpin(1,10):=Spin(1,10)\times \bbR^+$. The Frobenius tensor is defined as
follows: for any two odd vector fields $X,Y$ (i.e. sections of $F$) one
computes their Lie bracket and evaluates it modulo $F$. This defines, at each
point $p\in M$, a map $\wedge^2 F_p\rightarrow B_p=T_p/F_p$, and hence a
section of $\wedge^2 F^*\otimes B$. The requirement of $CSpin(1,10)$ invariance
means that one can choose bases $\{E_\a\},\  \{E^a\}$ for $F$ and $B^*$
respectively such that the components of the Frobenius tensor in such a basis,
denoted by $\hT_{\a\b}{}^c$, and by definition given by
\beq
\hT_{\a\b}{}^c=-\langle [E_\a, E_\b],E^c\rangle,
\eeq
where $\langle,\rangle$ denotes the usual pairing between vectors and forms,
are proportional to the Dirac matrices,
\beq \hT_{\a\b}{}^c=-i(\C^c)_{\a\b}.
\eeq
Note that, although we have used the usual torsion notation here, no connection
has been introduced as yet. Note also that the Dirac matrices define a tensor
which is $CSpin(1,10)$ invariant and not just $Spin(1,10)$ invariant.

For dimensions $D<11$ it is necessary to supplement this basic conformal
constraint with further constraints in order to obtain Poincar\'e supergavity,
but in $D=11$ this is not the case. In fact we can show that a special
superconformal structure on an $(11|32)$-dimensional superspace, $M$, is
equivalent to the equations of motion of eleven-dimensional supergravity,
modulo toplogical considerations.

To prove this one makes a choice of even tangent bundle $B$ (so that $T=F\oplus
B$) and introduces a suitable connection, $\hat \O$, which takes its values in
the Lie algebra of $CSpin(1,10)$. Thus one can write
\beqa
\hat\O_{\a}{}^{\b}&=&\O'_{\a}{}^{\b} +\d_{\a}{}^{\b} K\\
\hat\O_a{}^b&=&\O'_a{}^b +2\d_a{}^b K
\eeqa
where $K$ is the scale connection one-form, $\O'_{ab}=-\O'_{ba}$ is the
$\gs\gp\gi\gn(1,10)$ connection and
\beq
\O'_\a{}^\b={1\over4}(\C_{ab})_\a{}^\b \O'_{ab}
\eeq

The associated curvatures will be denoted $\hR$, $R'$ and $G$, the latter being
the scale curvature. The convention we shall use is that spinor (vector)
indices are lowered or raised with $\h_{\a\b}$ ($\h_{ab}$) and their inverses,
written with upper indices, where $\h_{\a\b}$ is the charge conjugation matrix
and is antisymmetric and $\h_{ab}$ is the usual Lorentz metric. These
operations are not covariant with respect to scale transformations so that one
has to keep track of the indices which have been raised or lowered in this way.
We shall also write the torsion with a hat in Weyl superspace as we shall later
transform back to Lorentzian superspace.

By a suitable choice of $B$ (in other words of $E_a$) and of the spinorial part
of the connection one can arrange that the dimension one-half components of the
torsion tensor, $\hT_{\a\b}{}^{\c}$ and $\hT_{\a b}{}^c$, vanish. At dimension
one we may choose the vectorial components of the connections such that
\beq
\hT_{ab}{}^c=0;\qquad (\C_a)^{\a\b} G_{\a\b}=0.
\eeq
The Bianchi identities then imply that \footnote{$H$ in this paper differs by a
sign to that of \cite{bh}.}
\beq
\hT_{a\b}{}^{\c}=-{1\over36}(\C^{bcd})_{\b}{}^{\c}H_{abcd}-{1\over288}
(\C_{abcde})_{\b}{}^\c H^{bcde}
\eeq
as well as
\beq
G_{\a\b}=0,
\eeq
where $H_{abcd}$ is totally antisymmetric and has Weyl weight 2. For
completeness we give the curvature tensor at dimension one:
\beq
R'_{\a\b,ab}={1\over6}\big((\C^{cd})_{\a\b}H_{abcd}+{1\over3}
(\C_{abcdef})_{\a\b} H^{cdef}\big).
\eeq

It is now a simple matter to prove that $G=0$ by using the Bianchi identity for
the scale curvature, $dG=0$. At dimension three-halves this identity gives,
since the dimension one component of $G$ vanishes,
\beq
\hT_{(\a\b}{}^c G_{\c)c}=0,
\eeq
which implies that $G_{\a b}=0$. At dimension two one then has
\beq
\hT_{\a\b}{}^c G_{cd}=0\Rightarrow G_{cd}=0.
\eeq
The analysis of the remaining dimension three-halves and two Bianchi identities
is then the same as in the original papers except that the covariant
derivatives  which arise include the scale connection $K$. At dimension
three-halves one finds that the dimension three-halves torsion is expressible
as a derivative of $H$:
\beq
\hT_{ab}{}^{\a}=-{i\over42}(\C^{cd})^{\a\b}\hD_{\b} H_{abcd}
\eeq
and that there are no further independent components of $H$ at this level,
\beq
\hD_{\a} H_{abcd}=-6i(\C_{[ab})_{\a}{}^{\b}\hT_{cd]\b}.
\eeq
The dimension three-halves curvature is given by
\beq
R'_{\a b,cd}=-{i\over2}\big(\C_b\hT_{cd}-\C_c\hT_{db}+\C_d\hT_{cb}\big)_{\a}
\eeq
where
\beq
(\C_b\hT_{cd})_{\a}:=(\C_b)_{\a}{}^{\b}\hT_{cd\b}.
\eeq
We also find the following torsion constraint
\beq
(\C^{abc})_{\a}{}^{\b}\hT_{bc\b}=0.
\eeq
The leading component of this equation is the field equation for the gravitino.

At dimension two one shows that the curvature tensor is expressed as a second
spinorial derivative of $H$ and that this is the only independent component of
$H$ at this level. One also finds the equations of motion for the graviton
\beq
R'_{ab}-{1\over2}\h_{ab}R'=-{1\over48}(4 H_{acde} H_b{}^{cde}
-{1\over2}\h_{ab}H_{cdef} H^{cdef})
\eeq
and
\beq
\hD^a H_{abcd}={1\over36.48}\e_{bcde_1\ldots e_8}H^{e_1\ldots e_4}H^{e_5\ldots
e_8}.
\label{heq}
\eeq

To complete the proof it is necessary to show that the four-index field $H$ can
be derived from a three-form potential, in which case (\ref{heq}) will be the
equation of motion for this field. Given the above results one may construct a
superspace four-form $H_4$ with Weyl weight -6 which is covariantly closed,
\beq
\hD H_4:=d H_4+6 H_4\wedge K=0
\label{hbi}
\eeq
and which has non-vanishing components $H_{abcd}$ and
\beq
H_{\a\b cd}=-i(\C_{cd})_{\a\b}.
\eeq
The proof that $H_4$ defined as above satisfies the Bianchi identity
(\ref{hbi}) is essentially the same as it is in the absence of the scale
connection, and will not be repeated here.
Since the curvature associated with $K$ vanishes we can deduce the existence of
a three-from potential  $B_3$, also of Weyl weight -6, such that
\beq
H=\hD B_3.
\eeq
Hence the proof is complete. Since the equations of motion are Weyl invariant
and since the scale curvature $G$ vanishes it follows that, if $H^1(M)=0$, $K$
can be transformed to zero so that we recover the standard equations of motion
of eleven-dimensional supergravity in superspace form.

As in the standard on-shell formalism it is possible to construct a seven-form
$H_7$  \cite{cl} which, in this case, has Weyl weight -12 and which satisfies
\beq
\hD H_7={1\over2}(H_4)^2.
\eeq
Its non-vanishing components are
\beqa
H_{\a\b abcde}&=&-i(\C_{abcde})_{\a\b}\\
H_{abcdefg}&=&{1\over4!}\e_{abcdefghijk} H^{hijk}.
\eeqa
It can be written in terms of potentials as
\beq
H_7=\hD B_6 + {1\over2} B_3\wedge H_4.
\eeq

We shall now rewrite the above equations in Lorentzian superspace by reducing
the structure group to $Spin(1,10)$. This can be accomplished by simply
taking the connection to be $\O'$ so that the Weyl connection $K$ now appears
in the torsion, that is,
\beqa
\hT_{AB}{}^c&=&T'_{AB}{}^c +2(K_A\d_B{}^c-(-1)^{AB}K_B\d_A{}^c) \nonumber\\
\hT_{AB}{}^{\c}&=&T'_{AB}{}^{\c} +(K_A\d_B{}^{\c}-(-1)^{AB}K_B\d_A{}^{\c})
\eeqa
where $T'$ is the torsion constructed using only $\O'$. However, this torsion
does not satisfy the standard constraint that its dimension one component with
purely vectorial indices vanish. In order to achieve this it is necessary to
make a further redefinition of the purely vectorial part of the connection. It
is simpler to combine this redefinition with the reduction of the structure
group in one step. We therefore set
\beqa
\hat \O_{\a \b}{}^{\c}&=&\O_{\a \b}{}^{\c}+ K_{\a}\d_{\b}{}^{\c}\nonumber\\
\hat \O_{\a b}{}^{c}  &=&\O_{\a b}{}^{c}+ 2K_{\a}\d_{b}{}^{c}\nonumber\\
\hat \O_{a \b}{}^{\c}&=&\O_{a \b}{}^{\c}+ K_{a}\d_{\b}{}^{\c} -
(\C_a{}^b)_{\b}{}^{\c}K_b\nonumber\\
\hat \O_{a b}{}^{c}  &=&\O_{a b}{}^{c}+ 2(K_{a}\d_{b}{}^{c}+\d_a{}^cK_b-
\h_{ab}K^c).
\eeqa
The components of the new Lorentzian torsion are, at dimension zero,
\beq
T_{\a\b}{}^c=-i(\C^c)_{\a\b},
\eeq
at dimension one-half,
\beqa
T_{\a b}{}^c=-2\d_b{}^c K_{\a};\qquad T_{\a\b}{}^{\c}=-2\d_{(\a}{}^{\c}
K_{\b)},
\eeqa
and, at dimension one,
\beq
T_{a\b}{}^{\c}=-{1\over36}\big((\C^{bcd})_{\b}{}^{\c}H_{abcd}+{1\over8}
(\C_{abcde})_{\b}{}^\c H^{bcde}\big) -(\C^b \C_a)_{\b}{}^{\c} K_b,
\eeq
as well as $T_{ab}{}^c=0$. The dimension three-halves torsion, since it does
not involve a connection, is unchanged. For the curvature one has
\beqa
R_{\a\b,cd}&=&R'_{\a\b,cd} +4iK_{[c}(\C_{d]})_{\a\b} \nonumber\\
R_{\a b,cd}&=&R'_{\a b,cd}-4\h_{b[c}D_{\a}K_{d]}+8\h_{b[c}K_{d]}\nonumber\\
R_{ab}{}^{cd}&=&R'_{ab}{}^{cd}+8\d_{[a}{}^{[c}D_{b]}K^{d]}-
16\d_{[a}{}^{[c}K_{b]}K^{d]}-8\d_{[a}{}^c\d_{b]}{}^d K^2,
\eeqa
where $R'$ is the curvature in the original superspace and $K^2:=K^a K_a$. From
the last of these equations one can compute the new Ricci tensor,
\beq
R_{ab}=R'_{ab}-18D_a K_b -2\h_{ab} D^c K_c +36(K_a K_b-\h_{ab} K^2).
\eeq
The one-form $K=E^{\a} K_{\a} +E^a K_a$ is of course still closed. If it is
exact it can be written as $dS$ for some $S$, and by making an appropriate
super-Weyl transformation all the terms involving $K$ can be removed leaving
the standard on-shell superspace again.

In summary, we have shown that the equations of motion of eleven-dimensional
supergravity are implied by the standard constraint on the dimension zero
torsion, at least if $M$ is simply connected. The proof of this assertion is
simplified by the use of Weyl superspace, although it is not essential to
introduce a scale connection. However, in Lorentzian superspace one has to
prove that $K$ is closed and recognise that it can be removed by a super-Weyl
transformation.

The formalism given here suggests a slight generalisation of standard
eleven-dimensional supergravity when $M$ is not simply connected. For example,
one might take $M$ to have the form $M^{10|32}\xz S^1$ with $K\sim mdy$, where
$y$ is the $S^1$ coordinate, thereby introducing a mass into the theory. This
possibility is discussed elsewhere \cite{hlw}.

We conclude with some brief comments on a recent paper \cite{ng} by Nishino and
Gates in which it is claimed that an off-shell extension of eleven-dimensional
supergravity can be constructed in superspace involving a dimension one-half
superfield. The authors use a Lorentzian structure group and include a closed
four-form $H$ (called $F$ in \cite{ng}) from the beginning. In their notation,
the basic dimension zero constraints are taken to be
\beq
T_{\a\b}{}^c=i(\C^c)_{\a\b};\qquad H_{\a\b cd}={1\over2}(\C_{cd})_{\a\b}
\eeq
The dimension one-half torsion components are
\beqa
T_{\a\b}{}^{\c}&=&-8(\C^a)_{\a\b}(\C_a)^{\c\d}J_{\d} \\
T_{\a b}{}^c&=& 8(\C^c\C_b)_{\a}{}^{\b} J_{\b}
\eeqa
and the dimension one-half component of $H$ is
\beq
H_{abc\d}=12i(\C_{abc})_{\d}{}^{\e} J_{\e}
\eeq
while the components of $H$ with negative dimensions, $H_{\a\b\c\d}$ and
$H_{\a\b\c d}$, are assumed to vanish. The field $J_{\a}$ is the new auxiliary
spinor superfield. However, it is not difficult to see that it can be removed
from the dimension one-half tensors by a field redefinition of the form
\beq
E_{a}\rightarrow E'_a=E_a +\L_a{}^{\a}E_{\a};\qquad E'_{\a}=E_{\a}
\eeq
which amounts to a change of choice of even tangent bundle. The connection form
is unchanged although the components of the vectorial part of the connection
will change as a consequence of the above change of basis. In terms of
differential forms one has
\beq
E'^a=E^a;\qquad E'^{\a}=E^{\a}-E^a\L_a{}^{\a}
\eeq
so that the torsion two forms change by
\beqa
T'^a&=& T^a \\
T'^{\a}&=& T^{\a}-T^a\L_a{}^{\a} -E^a D\L_a{}^{\a}
\eeqa
It is straightforward to compute the new torsion components as well as the
change in the components of $H$. At dimension zero (and less) there is no
change, and at dimension one-half one finds
\beqa
T'_{\a b}{}^c&=&T_{\a b}{}^c-\L_b{}^{\b} T_{\a\b}{}^c \\
T'_{\a\b}{}^{\c}&=&T_{\a\b}{}^{\c}-T_{\a\b}{}^c\L_c{}^{\c}\\
H'_{abc\d}&=& H_{abc\d} + 3\L_{[a}{}^{\c}H_{bc]\c\d}
\eeqa
If one makes the choice
\beq
\L_a{}^{\a}=8i(\C_a)^{\a\b} J_{\b}
\eeq
it is easy to see that the new dimension one-half components of the torsion and
$H$ are zero. There are more complicated changes induced at dimension one and
higher, but whatever they are we know from the results of \cite{h,cl}, or from
a simple adaptation of the argument given earlier in this paper, that the
equations of motion of eleven-dimensional supergravity will result. Hence the
formalism of Nishino and Gates is strictly equivalent to standard on-shell
supergravity in eleven dimensions. It differs from that of the current paper in
that, even if spacetime is topologically trivial, the transformations required
to recover standard on-shell superspace are not the same. In the Nishino-Gates
approach this is accomplished by the above redefinition of $E_a$ whereas in our
approach one needs to use a super-Weyl transformation. It is only after this
transformation has been made that one finds that $H$ is closed in the usual
sense.

A consequence of the above discussion is that the standard dimension zero
constraint on the torsion tensor needs to be amended in order to incorporate
the higher order corrections to supergravity which one would expect to arise in
$M$-theory.

\vskip .5cm
{\large{\bf Appendix}}

In this appendix we give a few more details of the algebraic steps involved in
the proof of the main result in the text. We suppress the hats throughout the
appendix as all quantities are taken to be in the Weyl superspace.

There are two main computations to perform, at dimension one-half and at
dimension one. At dimension one-half one is free to make redefinitions of the
form
\beq
E_a\rightarrow E_a +\L_a{}^{\a} E_{\a}
\la{a1}
\eeq
corresponding to making a choice of $B$, and one can impose further constraints
in order to solve algebraically for the dimension one-half part of the
connection. Using these freedoms, but without fixing the scale connection,
it is not difficult to see that one can bring $T_{\a b}{}^c$ to the form
\beq
T_{\a b,c}=\tilde T_{\a bc}
\la{a2}
\eeq
where the tensor on the right-hand side is symmetric and traceless on its
vector indices and gamma-traceless. This is an irreducible representation of
$Spin(1,10)$ which does not occur in the decomposition of the other dimension
one-half torsion $T_{\a\b}{}^{\c}$ into irreducibles. The dimension one-half
Bianchi identity is, when $T_{\a\b}{}^c=-i(\C^c)_{\a\b}$,
\beq
T_{(\a\b}{}^E T_{|E|\c)}{}^d=0.
\la{a3}
\eeq
This must be satisfied by the $T_{\a b}{}^c$ of (\ref{a2}), but it is not
difficult to see that this is not possible, and so we must have
\beq
T_{\a b}{}^c=0.
\la{a4}
\eeq
The Bianchi identity (\ref{a3}) then simplifies to
\beq
T_{\a\b}{}^{\e}(\C^d)_{\c)\e}=0
\la{a5}
\eeq
The remaining dimension one-half torsion is decomposed into irreducibles as
follows
\beq
T_{\a\b}{}^{\c}=\sum_{n=1,2,5}(\C^{a_1\ldots a_n})_{\a\b}\Psi_{a_1\ldots
a_n}^{\c}
\la{a6}
\eeq
with each $\Psi$ being a sum of  gamma-traceless antisymmetric tensor-spinors
with up to $n$ indices, for example,
\beq
\Psi_{ab}=\psi_{ab} +\C_{[a}\psi_{b]} +\C_{ab}\psi
\la{a7}
\eeq
where each of the $\psi'$s is irreducible. One thus has one each of such fields
with three, four or five indices, two fields with two indices and three fields
with zero and one index. Examining the Bianchi identity(\ref{a5}) for each of
these representations in turn one finds that they all vanish except for the
zero-index representations, of which only one is independent. The result of all
this is that, making use of the completeness relation for symmetric
gamma-matrices, one can write $T_{\a\b}{}^{\c}$ in the form
\beq
T_{\a\b}{}^{\c}=16\d_{(\a}{}^{\c}\L_{\b)}+
6(\C^a)_{\a\b}(\C_a)^{\c\d}\L_{\d}-(\C^{ab})_{\a\b}(\C_{ab})^{\c\d} \L_{\d}
\la{a8}
\eeq
This form for $T_{\a\b}{}^{\c}$ was derived in \cite{cl}, but there is a
difference here, namely that we still have the freedom to impose a constraint
corresponding to the dimension one-half scale connection. Using this freedom
one can impose $\L=0$ so that all dimension one-half torsion components vanish
as required. This final redefinition must be accompanied by further
redefinitions of the Lorentzian part of the connection and of $E_a$. The
required redefinition is
\beqa
\O_{\a,b}{}^c & \rightarrow & \O_{a,b}{}^c +(\C_b{}^c)_{\a}{}^{\b}\L_{\b}
+\d_b{}^c\L_{\a}\nonumber \\
E_a &\rightarrow & E_a +i(\C_a)^{\a\b}\L_{\a}E_{\b}
\la{a9}
\eeqa
Implementing this one finally arrives at
\beq
T_{\a\b}{}^{\c}=T_{\a b}{}^c=0.
\la{a10}
\eeq

We now turn to the dimension one analysis. There are two Bianchi identities:
\beq
R_{\a\b,cd} +2\h_{cd}G_{\a\b}=-2iT_{c(\a}{}^{\c}(\C_d)_{\b)\c}
\la{a11}
\eeq
and
\beq
R_{(\a\b,\c)}{}^{\d}+\d_{(\a}{}^{\d}G_{\b\c)}=-i(\C^c)_{(\a\b}T_{c\c)}{}^{\d}
\la{a12}
\eeq
The dimension one torsion may be written in the form
\beq
T_{a\b}{}^{\c}=\sum_{I=0}^{I=5}(\C^I)_{\b}{}^{\c}T_{a,I}
\la{a13}
\eeq
where the sum runs over the antisymmetrised $\C$-matrices up to rank five. Each
$T_{a,I}$ decomposes into three irreducible representations (except for $I=0$),
\beq
T_{a,b_1\ldots b_n}=T_{ab_1 \ldots b_n} +\h_{a[b_1}T_{b_2\ldots b_n]} +\tilde
T_{a,b_1\ldots b_n}
\la{a14}
\eeq
The first two of these are totally antisymmetric while the third is traceless
with vanishing totally antisymmetric part. $G_{\a\b}$ can be written
\beq
G_{\a\b}=(\C^{ab})_{\a\b} G_{ab} + (\C^{abcde})_{\a\b} G_{abcde}
\la{a15}
\eeq
since we may impose $(\C_a)^{\a\b}G_{\a\b}=0$ as a conventional constraint
corresponding to $K_a$. (Note that $G_{ab}$ here has dimension one and is not
the same as the dimension two component of the two-form $G$.) Substituting
these expressions in the symmetric part (on $c,d$) of the Bianchi identity
(\ref{a11}), one finds a number of constraints on the tensors $T_{a,I}$. It is
not difficult to verify that all of the $\tilde T$ tensors must vanish and that
there are further relations between the remaining antisymmetric tensors. The
result is
\beqa
T_a &=& 0  \nonumber \\
T_{a,b} &=& \h_{ab} T-2iG_{ab}\nonumber \\
T_{a,bc} &=& T_{abc}\nonumber \\
T_{a,bcd} &=& T_{abcd}+i\h_{a[b} G_{cd]}\nonumber \\
T_{a,bcde} &=& 5iG_{abcde}+\h_{a[b}T'_{cde]}\nonumber \\
T_{a,bcdef} &=& \h_{a[b}T'_{cdef]}\nonumber \\
T_{a,bcdefg}&=&-i\h_{a[b} G_{cdefg]}
\eeqa
where in the last line we have rewritten part of $T_{a,bcdef}$ in the
form $T_{a,bcdefg}$ by dualising on the last five indices. One may now use the
antisymmetric part of (\ref{a11}) to find the Lorentz curvature in terms of
these antisymmetric tensors. Finally one substitutes these expressions into the
Bianchi identity (\ref{a12}) and finds that they all vanish except those with
four indices and that there is only one independent one of these. We thus
arrive at the result (7) quoted in the text.

\vskip .5cm
{\bf Acknowledgement.} I thank Jim Gates for interesting discussions.

\end{document}